\documentclass[letterpaper,12pt]{article}

\usepackage{times}
\usepackage{fullpage}
\usepackage{amsmath}
\usepackage{amssymb}
\usepackage{amsthm}
\usepackage{amsfonts}
\usepackage{latexsym}

\newcommand \tinyspace {\mspace{1mu}}
\newcommand \norm[1] {\left\|\tinyspace#1\tinyspace\right\|}
\newcommand \tnorm[1] {\norm{#1}_{\mathrm{tr}}}
\newcommand \dnorm[1] {\norm{#1}_{\diamond}}

\newcommand{\ket}[1]{|\hspace{0.5pt}#1\hspace{0.5pt}\rangle}

\newtheorem{theorem}{Theorem}

\newtheorem{cor}[theorem]{Corollary}

\newtheorem{fact}[theorem]{Fact}
\theoremstyle{definition}

\newcommand{\cls}{\mathsf}
\newcommand{\reg}{\mathsf}
\newcommand{\mapset}{\mathbf}
\newcommand{\hilb}{\mathcal}

\DeclareMathOperator{\tr}{tr}
\DeclareMathOperator{\dist}{dist}

\begin{document}

\title{Quantum Interactive Proofs with Competing Provers}

\author{Gus Gutoski \hspace{1cm} John Watrous\\[3mm]
Department of Computer Science\\
University of Calgary\\
Calgary, Alberta, Canada}

\date{December 13, 2004}

\maketitle

\begin{abstract}
This paper studies quantum refereed games, which are quantum interactive proof
systems with two competing provers: one that tries to convince the verifier to
accept and the other that tries to convince the verifier to reject.
We prove that every language having an ordinary quantum interactive proof
system also has a quantum refereed game in which the verifier exchanges just
one round of messages with each prover.
A key part of our proof is the fact that there exists a single quantum
measurement that reliably distinguishes between mixed states chosen
arbitrarily from disjoint convex sets having large minimal trace distance from
one another.
We also show how to reduce the probability of error for some classes of
quantum refereed games.
\end{abstract}

%=============================================================================%

\section{Introduction} \label{sec:intro}

A \emph{refereed game} consists of a conversation between a computationally
bounded verifier and two computationally unbounded provers regarding some
input string $x$.
The two provers use their unbounded computational power to compete with each
other: one prover, called the \emph{yes-prover}, attempts to convince the
verifier to accept $x$, while the other prover, called the \emph{no-prover},
attempts to convince the verifier to reject $x$.
At the end of the interaction, the verifier decides whether to accept or
reject the input $x$, effectively deciding which of the provers wins the game.
Such games represent games of incomplete information; the messages exchanged
between one prover and the verifier are considered to be hidden from the other
player.

A language $L$ is said to have a refereed game with error $\varepsilon$ if
there is a polynomial-time verifier satisfying the following conditions.
For each string $x \in L$, there exists a yes-prover that can always convince
the verifier to accept $x$ with probability at least $1-\varepsilon$,
regardless of the no-prover's strategy, and for each $x \not \in L$, there
exists a no-prover that can always convince the verifier to reject $x$ with
probability at least $1 - \varepsilon$, regardless of the yes-prover's
strategy.
A \emph{turn} for one of the provers consists of a message from the verifier
to that prover, followed by a response from that prover back to the verifier.
One may consider the case where the provers' turns are played sequentially
or in parallel.

The refereed games model is based on the interactive proof system model
\cite{GoldwasserM+89,Babai85,BabaiM88,Ben-OrG+88}, which has a rich
history that we will not survey here.
The refereed games model, and variations on this model, were considered in
the classical case in
Refs.~\cite{Reif84,FeigeS+90,FeigeS92,KollerM92,FeigenbaumK+95,FeigeK97},
among others.
Much of what is known about the complexity-theoretic aspects of the classical
refereed games model is due to Feige and Kilian~\cite{FeigeK97}.
The class of languages having classical refereed games in which the provers
may play any polynomial number of turns coincides with $\cls{EXP}$
(deterministic time $2^{p(n)}$ for some polynomial $p$).
The simulation of $\cls{EXP}$ by a polynomial-turn refereed game is due to
Feige and Kilian~\cite{FeigeK97}, and is based on arithmetization technique
developed by Lund, Fortnow, Karloff and Nisan \cite{LundF+92} and used in
proofs of $\cls{IP} = \cls{PSPACE}$ \cite{Shamir92,Shen92}.
The simulation of polynomial-turn refereed games in $\cls{EXP}$ is due to
Koller and Megiddo~\cite{KollerM92}.
On the other hand, the class of languages having games in which the provers
play precisely one turn each, with the turns played in parallel, coincides
with $\cls{PSPACE}$ \cite{FeigeK97}.
Apparently little is known about the expressive power of classical refereed
games intermediate between these two extremes.
For instance, games with a constant number of prover turns may correspond to
$\cls{PSPACE}$, $\cls{EXP}$, or some complexity class between the two.

Similar to the classical case, quantum refereed games are based on the
quantum interactive proof system model \cite{Watrous03,KitaevW00}.
Quantum refereed games differ from classical ones in that the provers
and the verifier may perform quantum computations and exchange quantum
messages.
Our two main motives for considering the quantum refereed games model
are to better understand the power of quantum interactive proof systems
and to examine the effect of quantum information on the complexity of
finding strategies for two-player games.

The main result of this paper establishes that any language having a quantum
interactive proof system also has a quantum refereed game with exponentially
small probability of error wherein each prover plays just one turn (with the
yes-prover playing first).
An interesting fact about the resulting game from the point of view of
understanding quantum interactive proofs is that entanglement between
the provers and the verifier does not play any role in this game.
More specifically, the game we define has the following general form:
the yes-prover sends the verifier a mixed quantum state, the verifier
processes this state and sends some state to the no-prover, and the no-prover
measures the state and sends a classical result to the verifier.
The verifier checks the result of the measurement and accepts or rejects.

A key ingredient for our result is an information-theoretic assertion stating
that there exists a quantum measurement that can reliably distinguish between
states chosen from two disjoint convex sets of quantum states.
This assertion generalizes a well-known fact about the relation between
the trace distance between two states and their distinguishability, and
may be viewed as a quantitative version, from the point of view of quantum
information theory, of the fact from convex analysis that disjoint convex
sets are separated by some hyperplane.

The remainder of this paper is organized as follows.
We begin by defining quantum refereed games in Section~\ref{sec:defs}.
In Section~\ref{sec:qsep} we prove the fact concerning measurements
distinguishing convex sets mentioned previously.
Using this fact, we then prove in Section~\ref{sec:QIPinQRG} that a two-turn
quantum refereed game exists for any language having a quantum interactive
proof system.
In Section~\ref{sec:error} we describe a method for error reduction in two-turn
quantum refereed games.
The paper concludes with Section~\ref{sec:conclusion}, which mentions some
open problems.

%=============================================================================%

\section{Definitions} \label{sec:defs}

In this section we define the quantum refereed games model and some complexity
classes based on this model.
Throughout the paper we assume all strings are over the alphabet
$\Sigma = \{0,1\}$.
For $x \in \Sigma^*$, $|x|$ denotes the length of $x$.
We let $\mathit{poly}$ denote the set of polynomial-time computable functions
$f : \mathbb{N} \to \mathbb{N} \setminus \{0\}$ for which there exists a
polynomial $p$ such that $f(n) \leq p(n)$ for all $n$.
We also let $2^{-\mathit{poly}}$ denote the set of polynomial-time computable
functions $\varepsilon$ such that $\varepsilon(n) = 2^{-f(n)}$ for all $n$
for some $f\in\mathit{poly}$.

The model for quantum computation that provides a basis for quantum refereed
games is the quantum circuit model, with which we assume the reader is
familiar.
As mentioned in Section~\ref{sec:intro}, a quantum refereed game has a verifier
$V$ and two competing provers $Y$ and $N$.
Each of $V$, $Y$, and $N$ is defined
by a mapping on input strings $x \in \Sigma^*$ where $V(x)$, $Y(x)$, and $N(x)$
are each sequences of quantum circuits.
The circuits in these sequences are assumed to be composed only of gates
taken from some universal set of quantum gates.
Thus, each of the circuits implements a unitary operation on its input qubits.
However, we lose no generality by allowing only unitary operations because
arbitrary admissible quantum operations, including measurements, can be
simulated by unitary circuits as described in Ref.~\cite{AharonovK+98}.

For each prover, the qubits upon which that prover's circuits act are
partitioned into two sets: one set of qubits is private to that prover and
the other is shared with the verifier.
These shared qubits act as a quantum channel between the verifier and that
prover.
No restrictions are placed on the complexity of the provers' circuits,
which captures the notion that the provers' computational power is
unbounded---each of the provers' circuits can be viewed as an arbitrary
unitary operation.

The qubits on which the verifier's circuits act are partitioned into three
sets: one set is private to the verifier and two sets are shared with each of
the provers.
One of the verifier's private qubits is designated as the \emph{output qubit}.
At the end of the game, acceptance is dictated by a measurement of the output
qubit in the computational basis.
We also require that the verifier's sequence of circuits $V(x)$ be generated by
a polynomial-time Turing machine on input $x$.
This uniformity constraint captures the notion that the verifier's
computational power is limited.

In addition to the verifier and provers, a quantum refereed game consists of
a \emph{protocol} that dictates the number and order of turns taken by the
provers.
The circuits in the verifier's and provers' sequences are applied to the
initial state in which each qubit is in state $\ket{0}$ in such a way as to
implement the protocol of the game.

The games we study in this paper have the following protocol: a message from
the yes-prover to the verifier, a message from the verifier to the no-prover,
and a message from the no-prover the the verifier.
Quantum refereed games that follow this protocol will be called
\emph{short quantum games}.
We note that entanglement between the provers and the verifier is immaterial
in games of this form---each prover takes only one turn, and thus has no need
to remember anything after his turn ends.
Thus, when convenient, we may assume that the provers do not have private
qubits but instead may perform arbitrary admissible quantum operations
(i.e., completely positive trace-preserving maps) on their message qubits.

We now define the complexity class $\cls{SQG}$ based on short quantum games
of the type just described.
For $c,s : \mathbb{N} \to [0,1]$, the set $\cls{SQG}(c,s)$ consists of all
languages $L\subseteq \Sigma^*$ for which there exists a verifier $V$ for a
short quantum game such that the following conditions hold:
\begin{enumerate}
\item
There exists a yes-prover $Y$ such that, for all no-provers $N$ and all
$x \in L$, $Y(x)$ convinces $V(x)$ to accept $x$ with probability at least
$1 - c(|x|)$; and
\item
There exists a no-prover $N$ such that, for all yes-provers $Y$ and all
$x\not\in L$, $N(x)$ convinces $V(x)$ to reject $x$ with probability at
least $1 - s(|x|)$.
\end{enumerate}
The functions $c$ and $s$ are called the {\em completeness error} and
{\em soundness error}, respectively.
We write $\cls{SQG}$ to denote the class of all languages
$L \subseteq \Sigma^*$ such that $L \in \cls{SQG}(\varepsilon,\varepsilon)$
for every $\varepsilon \in 2^{-\mathit{poly}}$.

The class $\cls{QIP}$ contains all problems having single-prover quantum
interactive proof systems as in Ref.~\cite{KitaevW00}.
The main complexity-theoretic result of the present paper states that
$\cls{QIP} \subseteq \cls{SQG}$.
We prove this result by exhibiting a short quantum game that solves a promise
problem called the \textsc{close-images} problem, which is known to be complete
for $\cls{QIP}$ \cite{KitaevW00}.
It is convenient for us to use the formulation of this problem based on the
one found in Ref.~\cite{RosgenW04}.

The promise problem \textsc{close-images} is defined for any desired
$\varepsilon \in 2^{-\mathit{poly}}$ as follows.
Given are descriptions of two mixed-state quantum circuits $Q_0$ and $Q_1$,
which both implement some admissible transformation from $n$ qubits to $m$
qubits.
The promise is that exactly one of the following conditions holds:
\begin{enumerate}
\item
There exist $n$-qubit mixed states $\rho_0$ and $\rho_1$ such that
$Q_0(\rho_0) = Q_1(\rho_1)$; or
\item
For all $n$-qubit mixed states $\rho_0$ and $\rho_1$, the states $Q_0(\rho_0)$
and $Q_1(\rho_1)$ have fidelity squared at most $\varepsilon(n)$.
\end{enumerate}
In other words, the images of $Q_0$ and $Q_1$ are either overlapping or are
far apart.
The goal is to accept when case 1 holds and reject when case 2 holds.

%=============================================================================%

\section{Distinguishing Convex Sets of Quantum States} \label{sec:qsep}

We motivate discussion in this section by pointing out that, for any
mixed-state quantum circuit $Q$, the image
$\mathcal{A} = \{Q(\rho):\rho \textrm{ a mixed state}\}$ of the admissible
transformation associated with $Q$ is a compact, convex set of mixed states.
If the images of two circuits $Q_0$ and $Q_1$ are far apart, then one
could reasonably hope that there is a quantum measurement that reliably
distinguishes between outputs $Q_0(\rho_0)$ and $Q_1(\rho_1)$ of these
transformations, with the measurement depending only on $Q_0$ and $Q_1$,
and not on the choice of input states $\rho_0$ and $\rho_1$.
In this section we prove that indeed there always exists such a measurement.
More generally, we prove that given any two disjoint convex sets of mixed
quantum states, there exists a single measurement that distinguishes states
drawn arbitrarily from one set from the other with success probability
determined by the minimal trace distance between the sets.
The short quantum game for the \textsc{close-images} problem we define
in Section~\ref{sec:QIPinQRG} relies upon the existence of such a measurement.

Let us first begin with some notation.
Given a finite dimensional Hilbert space $\hilb{H}$, let
$\mapset{L}(\hilb{H})$ denote the set of all linear operators on $\hilb{H}$,
let $\mapset{H}(\hilb{H})$ denote the set of all Hermitian operators on
$\hilb{H}$, let $\mapset{Pos}(\hilb{H})$ denote the set of all positive
semidefinite operators on $\hilb{H}$, and let $\mapset{D}(\hilb{H})$ denote
the set of all density operators (i.e., unit trace positive semidefinite
operators) on $\hilb{H}$.
For $A,B\in\mapset{L}(\hilb{H})$, define
$\langle A,B\rangle = \tr A^{\dagger}B$.
This is an inner product on $\mapset{L}(\hilb{H})$ that is sometimes
called the Hilbert-Schmidt inner product.
%Note that if $A,B\in\mapset{H}(\hilb{H})$, then $\langle A, B\rangle$ is
%a real number.

For a vector $\ket{\psi}\in\hilb{H}$, $\norm{\ket{\psi}}$ denotes the
Euclidean norm of $\ket{\psi}$.
For an operator $A \in \mapset{L}(\hilb{H})$, the operator norm of $A$, denoted
$\norm{A}$, is defined by
\[
\norm{A} = \sup_{\ket{\psi} \in \hilb{H} \setminus \{0\} }
\frac{ \norm{A \ket{\psi}} }{ \norm{\ket{\psi}} }.
\]
The trace norm of $A$, denoted $\tnorm{A}$, is defined by
$\tnorm{A} = \tr\sqrt{A^{\dagger}A}$.
The trace norm and the operator norm are dual to one another with respect
to the Hilbert-Schmidt inner product, meaning that the following fact holds.
\begin{fact} \label{fact:trDual}
For every $A \in \mapset{L}(\hilb{H})$,
\begin{eqnarray*}
\norm{A} &=&
\max \left\{ |\langle B,A\rangle| : B \in \mapset{L}(\hilb{H}),\;
\tnorm{B} \leq 1 \right\},\\
\tnorm{A} &=&
\max \left\{ |\langle B,A\rangle| : B \in \mapset{L}(\hilb{H}),\;
\norm{B} \leq 1 \right\}.
\end{eqnarray*}
\end{fact}

\noindent
See, for instance, Bhatia~\cite{Bhatia97} for a proof of this fact.

The trace norm characterizes the distinguishability of a given pair of
density matrices $\rho_0,\rho_1\in\mapset{D}(\hilb{H})$ in the following
sense.
There exists a binary-valued quantum measurement such that if
$\rho\in \{\rho_0, \rho_1\}$ is chosen uniformly at random, then the
measurement correctly determines which of $\rho_0$ or $\rho_1$ was given
with probability $\frac{1}{2} + \frac{1}{4}\|\rho_0 - \rho_1\|_{\tr}$.
Furthermore, such a measurement is optimal in the sense that no other quantum
measurement can possibly distinguish between $\rho_0$ and $\rho_1$ with a
higher success rate.
An immediate corollary of this fact is that for a given pair
$\rho_0$ and $\rho_1$, there exists a measurement that correctly identifies
a chosen state $\rho\in\{\rho_0,\rho_1\}$ with probability of
correctness at least $\frac{1}{2}\tnorm{\rho_0 - \rho_1}$, even if $\rho$ is
chosen by an adversary that knows the measurement.

Consider the following variant of the distinguishability problem:
We are given $\rho \in \mapset{D}(\hilb{H})$ chosen from one of two disjoint
convex sets of density operators
$\mathcal{A}_0, \mathcal{A}_1 \subseteq \mapset{D}(\hilb{H})$, and we are
asked to determine the set from which $\rho$ was chosen.
For simplicity we will assume $\mathcal{A}_0$ and $\mathcal{A}_1$ are closed
sets.
Under this assumption, it is meaningful to define the trace distance
$\dist(\mathcal{A}_0,\mathcal{A}_1)$ between $\mathcal{A}_0$ and
$\mathcal{A}_1$ as the minimum of the quantity $\tnorm{\rho_0 - \rho_1}$ over
all choices of $\rho_0 \in \mathcal{A}_0$ and $\rho_1\in \mathcal{A}_1$.
We prove that there exists a single measurement with the property that
if an arbitrary $\rho$ is chosen from $\mathcal{A}_0$ with probability 1/2,
and otherwise $\rho$ is chosen from $\mathcal{A}_1$, then the measurement
correctly determines which set $\rho$ was chosen from with probability
at least $\frac{1}{2} + \frac{1}{4}\dist(\mathcal{A}_0,\mathcal{A}_1)$.
This fact therefore generalizes the fact concerning a single pair of
quantum states mentioned above, as singleton sets are of course closed
and convex.
As above, this fact implies that if $\rho$ is chosen from
$\mathcal{A}_0\cup\mathcal{A}_1$ in an arbitrary manner, even depending
on the measurement itself, then the measurement will correctly determine
from which of $\mathcal{A}_0$ or $\mathcal{A}_1$ the state $\rho$ was
chosen with probability at least
$\frac{1}{2}\dist(\mathcal{A}_0,\mathcal{A}_1)$.

The proof of this fact begins with a well-known result from convex analysis,
which informally states that there exists a separating hyperplane between any
two disjoint convex sets.
Typically, the separation result is stated in terms of the vector space
$\mathbb{R}^n$, but it translates to $\mapset{H}(\hilb{H})$ for a given
space $\hilb{H}$ without complications, as $\mapset{H}(\hilb{H})$ may
be identified with the vector space $\mathbb{R}^{m^2}$, for
$m = \dim(\hilb{H})$.
Here we state a restricted variant of this fact that is most convenient for our
purposes---see Rockafellar~\cite{Rockafellar70}, for instance, for a more
general statement.

\begin{fact} \label{fact:hyperplane}
Let $\mathcal{A},\mathcal{B} \subseteq \mapset{H}(\hilb{H})$ be disjoint
convex sets with $\mathcal{A}$ compact and $\mathcal{B}$ open.
Then there exists a Hermitian operator $H \in \mapset{H}(\hilb{H})$
and a real number $a\in\mathbb{R}$ such that
$\langle H, X\rangle  \geq a > \langle H, Y\rangle$ for all
$X\in \mathcal{A}$ and $Y\in\mathcal{B}$.
\end{fact}

We are now ready to state and prove the main result of this section.

\begin{theorem} \label{theorem:separate}
Let $\mathcal{A}_0, \mathcal{A}_1 \subseteq \mapset{D}(\hilb{H})$ be
closed convex sets of density operators.
Then there exist measurement operators $E_0,E_1\in\mapset{Pos}(\hilb{H})$
with $E_0 + E_1 = I$ such that the following holds.
For every pair $\rho_0\in\mathcal{A}_0$ and $\rho_1\in\mathcal{A}_1$,
if $\rho$ is chosen uniformly from $\{\rho_0,\rho_1\}$ and measured
via the measurement $\{E_0,E_1\}$, the measurement will correctly
determine whether $\rho\in\mathcal{A}_0$ or $\rho\in\mathcal{A}_1$
with probability at least
$\frac{1}{2} + \frac{1}{4}\dist(\mathcal{A}_0,\mathcal{A}_1)$.
\end{theorem}

\begin{proof}
Let $d = \dist(\mathcal{A}_0,\mathcal{A}_1)$.
If $d = 0$, the theorem is trivially satisfied by the measurement defined
by $E_0 = E_1 = \frac{1}{2}I$ (which is equivalent to a random coin-flip),
so assume that $d > 0$.
Let
\[
\mathcal{A} = \mathcal{A}_0 - \mathcal{A}_1 =
\{ \rho_0 - \rho_1 : \rho_0 \in \mathcal{A}_0, \rho_1 \in \mathcal{A}_1
\}.
\]
Then $\mathcal{A}$ is a compact convex set of Hermitian operators and
$\tnorm{X} \geq d$ for every $X \in \mathcal{A}$.
Let
\[ \mathcal{B} = \{ Y \in \mapset{H}(\hilb{H}) : \tnorm{Y} < d\} \]
denote the open ball of radius $d$ in $\mapset{H}(\hilb{H})$ with respect
to the trace norm.
The sets $\mathcal{A}$ and $\mathcal{B}$ satisfy the conditions of
Fact~\ref{fact:hyperplane},
and therefore there exists a Hermitian operator
$H \in \mapset{H}(\hilb{H})$ and a real number $a\in\mathbb{R}$ such that
$\langle H, X\rangle \geq a > \langle H, Y\rangle$ for all $X\in\mathcal{A}$
and $Y\in\mathcal{B}$.
Because $Y\in \mathcal{B}$ if and only if $-Y\in\mathcal{B}$ for every $Y$,
it follows that $-a < a$, and therefore $a > 0$.

Let $K = \frac{d}{a}H$.
We therefore have that $\langle K , X\rangle \geq d$ for every
$X\in\mathcal{A}$ and $\langle K, \frac{1}{d}Y\rangle < 1$ for every
$Y \in \mathcal{B}$.
As $\frac{1}{d}Y$ ranges over all Hermitian operators with trace norm smaller
than 1, this implies $\|K\| \leq 1$ by Fact~\ref{fact:trDual}.
Now, let $K^{+}, K^{-}\in\mapset{Pos}(\hilb{H})$ denote the positive
and negative parts of $K$, meaning that they satisfy
$K = K^{+} - K^{-}$ and $\langle K^+ , K^-\rangle = 0$.
As $\|K\|\leq 1$ it follows that $K^{+} + K^{-} \leq I$.

At this point we define $E_0,E_1 \in \mapset{Pos}(\hilb{H})$ as follows:
\[
E_0 = K^{+} + \frac{1}{2}(I - K^{+} - K^{-})\;\;\;\mbox{and}\;\;\;
E_1 = K^{-} + \frac{1}{2}(I - K^{+} - K^{-}) .
\]
The operators $E_0$ and $E_1$ are both positive semidefinite and satisfy
$E_0 + E_1 = I$, and therefore represent a binary-valued POVM.

Now suppose $\rho_0\in\mathcal{A}_0$ and $\rho_1\in\mathcal{A}_1$ are
chosen arbitrarily, and $\rho$ is chosen uniformly from the set
$\{\rho_0,\rho_1\}$.
Let $C$ denote the event that the measurement $\{E_0,E_1\}$ correctly
determines which of $\rho_0$ and $\rho_1$ was selected.
We have $\Pr[C] = \frac{1}{2}\langle E_0,\rho_0\rangle +
\frac{1}{2}\langle E_1,\rho_1\rangle$, and therefore
\[
\Pr[C] - \Pr[\lnot C] = \frac{1}{2}\langle E_0 - E_1, \rho_0 - \rho_1\rangle
= \frac{1}{2}\langle K,\rho_0 - \rho_1\rangle \geq \frac{d}{2}  ,
\]
with the inequality following from the fact that
$\rho_0 - \rho_1\in\mathcal{A}$.
Consequently the measurement is correct with probability at least
$\frac{1}{2} + \frac{d}{4}$ as required.
\end{proof}

%=============================================================================%

\section{A Short Quantum Game for $\cls{QIP}$}
\label{sec:QIPinQRG}

In this section, we prove that any language with a quantum interactive proof
system also has a short quantum game by solving the $\cls{QIP}$-complete
problem $\textsc{close-images}$ from Section~\ref{sec:defs}.

First, let us recall that the fidelity $F(\rho, \xi)$ between
two quantum states $\rho, \xi \in \mapset{D}(\hilb{H})$ is defined as
$F(\rho,\xi) = \left\| \sqrt{\rho} \sqrt{\xi} \right\|_{\mathrm{tr}}$.
The following fact, proved by Fuchs and van de Graaf~\cite{FuchsvdG99},
gives one relationship between the fidelity and the trace norm.

\begin{fact}
\label{fact:trF}
Let $\rho, \xi \in \mapset{D}(\hilb{H})$.
Then
\[
1 - \frac{1}{2}\| \rho - \xi \|_{\tr} \leq F(\rho,\xi) \leq
\sqrt{1 - \frac{1}{4} \| \rho - \xi \|_{\tr}}  .
\]
\end{fact}
We are now ready to state and prove the main result of this section.

\begin{theorem} \label{theorem:THEtheorem}
$\cls{QIP} \subseteq \cls{SQG} \left( 1/2,2^{-\mathit{poly}} \right)$.
\end{theorem}

\begin{proof}

It suffices to show that $\textsc{close-images}$ is in
$\cls{SQG}(1/2,2^{-\mathit{poly}})$.
Suppose the input encodes mixed state quantum circuits $Q_0$ and $Q_1$,
each mapping $n$ qubits to $m$ qubits.
Let $\hilb{H}$ and $\hilb{K}$ be Hilbert spaces with dimensions $2^n$ and
$2^m$ corresponding to the $n$ input qubits and $m$ output qubits respectively.
We may view $Q_0$ and $Q_1$ as corresponding to admissible transformations
$Q_0,Q_1 : \mapset{D}(\hilb{H}) \to \mapset{D}(\hilb{K})$.
Let $\mathcal{A}_i = \{Q_i(\rho) : \rho \in \mapset{D}(\hilb{H}) \} \subseteq
\mapset{D}(\hilb{K})$
denote the image of $Q_i$ for $i=0,1$.
The sets $\mathcal{A}_0$ and $\mathcal{A}_1$ are closed, convex sets of density
operators.

Consider the following verifier for a short quantum game:
\begin{enumerate}
\item Receive $n$-qubit registers $\reg{X}_0$ and $\reg{X}_1$
from the yes-prover.
\item Choose $i\in\{0,1\}$ uniformly at random and apply $Q_i$ to register
$\reg{X}_i$.
Let the output be contained in an $m$-qubit register $\reg{Y}$,
which is then sent to the no-prover.
\item Receive a classical bit $b$ from the no-prover.
Accept if $b\not=i$ and reject if $b=i$.
\end{enumerate}

If $(Q_0,Q_1)$ is a ``yes'' instance of \textsc{close-images} then there
exist $\rho_0,\rho_1 \in \mapset{D}(\hilb{H})$ such that
$Q_0(\rho_0) = Q_1(\rho_1)$.
The strategy for the yes-prover is to prepare the registers $\reg{X}_0$
and $\reg{X}_1$ in states $\rho_0$ and $\rho_1$, respectively, and to
send them to the verifier in step 1 of the verifier's protocol.
Because $Q_0(\rho_0) = Q_1(\rho_1)$, the state contained in the register
$\reg{Y}$ is independent of $i$, so the no-prover can do no better than
randomly guessing in step 3.
The verifier will therefore accept with probability $1/2$ in this case.

If $(Q_0,Q_1)$ is a ``no'' instance of \textsc{close-images} then for any
desired $\varepsilon \in 2^{-\mathit{poly}}$ we are promised that
\[
\sqrt{\varepsilon(n)}
\geq \max_{\xi_0,\xi_1 \in \mapset{D}(\hilb{H})}
\left\{ F(Q_0(\xi_0), Q_1(\xi_1)) \right\}
\geq
1 - \frac{1}{2} \dist(\mathcal{A}_0,\mathcal{A}_1)  .
\]
It follows that $\dist(\mathcal{A}_0,\mathcal{A}_1) \geq
2 - 2\sqrt{\varepsilon(n)}$.

Regardless of the state of the registers $\reg{X}_0$ and $\reg{X}_1$
sent to the verifier by the yes-prover, we must have that the reduced
state of the register $\reg{Y}$ sent to the no-prover is given by
some state $\xi \in \mathcal{A}_0 \cup \mathcal{A}_1$, and moreover that
$\Pr[\xi \in \mathcal{A}_0] = \Pr[\xi \in \mathcal{A}_1] = 1/2$.
By Theorem \ref{theorem:separate} there exists a quantum measurement
$\{E_0,E_1\}$ that correctly determines whether $\rho \in \mathcal{A}_0$ or
$\rho \in \mathcal{A}_1$ with probability at least
\[
\frac{1}{2} + \frac{1}{4}\dist(\mathcal{A}_0,\mathcal{A}_1)
\geq 1-\frac{\sqrt{\varepsilon(n)}}{2}  .
\]
The strategy for the no-prover is to perform the quantum measurement
$\{E_0,E_1\}$ and send the result to the verifier in step 3.
This causes the verifier to reject with probability at least
$1-\sqrt{\varepsilon(n)}/2$.
As this argument holds for every $\varepsilon\in 2^{-\mathit{poly}}$,
we have that the soundness error is $2^{-\mathit{poly}}$ as required.
\end{proof}

%=============================================================================%

\section{Error Reduction} \label{sec:error}

Suppose that both the completeness and soundness error $c$ and $s$ of a
refereed game are bounded below $1/2$ by an inverse polynomial.
Then it follows from Chernoff bounds that these error probabilities can be made
exponentially close to zero by repeating the game a polynomial number of times
in succession and taking a majority vote.
Of course, sequential repetition necessarily increases the number of turns in
the game and so it is natural to ask if error reduction can be achieved without
affecting the turn complexity of the game.

A natural approach to this task is to run many copies of the refereed game in
parallel and to accept or reject based on the outcomes of the repetitions.
This technique is purely classical and has been successfully applied to
classical single- and multi-prover interactive proof systems (see for example
Ref.~\cite{Raz98} and the references therein).
A potential problem with this technique is that the provers need not treat
each repetition independently---they might try to correlate the parallel
repetitions (or entangle them in the quantum case) in some devious way such
that the completeness and/or soundness error does not decrease as desired.

In the quantum setting, the general case of this problem has not been
completely solved.
But for three-message single-prover quantum interactive proof systems with
zero completeness error, Ref.~\cite{KitaevW00} proves that parallel
repetition followed by a unanimous vote does indeed achieve the exponential
reduction in soundness error that one might expect, regardless of any possible
entanglement by the prover among the parallel copies.

In this section, we prove that parallel repetition followed by a unanimous
vote can be used to improve the error bounds for short quantum games by
reducing the problem to error reduction for single-prover quantum interactive
proof systems with three or fewer messages.
The reduction is achieved by fixing a yes- or no-prover $P$ that is guaranteed
to win with a certain probability.
By viewing the verifier-prover pair $(V,P)$ as a new composite verifier, we
are left with what is now effectively a one- or two-message quantum
interactive proof system in which the opposing prover is the lone prover.
We define a verifier-prover pair $(V',P')$ that runs many copies of $(V,P)$
in parallel and accepts based on a unanimous vote.
We can then employ the error reduction result of Ref.~\cite{KitaevW00}
to prove that the error of the new game decreases exponentially in the number
of repetitions.

We formalize this argument shortly, but first we require additional notation.
Given finite-dimensional Hilbert spaces $\hilb{H}$ and $\hilb{K}$, let
$\mapset{L}(\hilb{H}, \hilb{K})$ denote the set of all linear operators
mapping $\hilb{H}$ to $\hilb{K}$ and let $\mapset{T}(\hilb{H}, \hilb{K})$
denote the set of all linear operators mapping the vector space
$\mapset{L}(\hilb{H})$ to $\mapset{L}(\hilb{K})$.
The trace norm can be extended to $\mapset{T}(\hilb{H},\hilb{K})$ as follows.
For $T \in \mapset{T}(\hilb{H},\hilb{K})$,
\[
\tnorm{T} = \sup_{X \in \mapset{L}(\hilb{H}) \setminus \{ 0 \} }
\frac{\tnorm{T(X)}}{\tnorm{X}}  .
\]
Let $\hilb{L}$ be a Hilbert space with $\dim(\hilb{L}) = \dim(\hilb{H})$ and
let $I_{\mapset{L}(\hilb{L})}$ denote the identity transformation on
$\mapset{L}(\hilb{L})$.
Then for $T \in \mapset{T}(\hilb{H},\hilb{K})$, the \emph{diamond norm}
$\dnorm{T}$ of $T$ is given by
$\dnorm{T} = \tnorm{ T \otimes I_{\mapset{L}(\hilb{L})} }$.
Further information on the diamond norm may be found in
Kitaev, Shen, and Vyalyi~\cite{KitaevS+02}.
The diamond norm satisfies several nice properties that the trace norm
(extended to $\mapset{T}(\hilb{H},\hilb{K})$) does not.
The diamond norm is multiplicative with respect to tensor products:
$\dnorm{T_1 \otimes T_2} = \dnorm{T_1} \dnorm{T_2}$ for any choice of
transformations $T_1$ and $T_2$.

We are now prepared to give the main result of this section, whose proof is
based on the proof of Theorem 6 of Ref.~\cite{KitaevW00}.

\begin{theorem} \label{theorem:boundedError}
$\cls{SQG}(c,s) \ \subseteq \ \cls{SQG} ( kc, s^k ) \cap \cls{SQG} (c^k, ks)$
for any choice of $c,s : \mathbb{N} \to [0,1]$ and $k \in \mathit{poly}$.
\end{theorem}

\begin{proof}
We first prove that $\cls{SQG}(c,s) \subseteq \cls{SQG} ( kc, s^k ).$
Let $L \in \cls{SQG}(c,s)$ and let $V(x)=(V(x)_1$, $V(x)_2)$ be a verifier
witnessing this fact.
For the remainder of this proof, we assume that the input $x \in \Sigma^*$ is
fixed.
For brevity we drop the argument and write $V=(V_1,V_2)$ and use similar
notation for the provers.

Let $V'=(V_1^{\otimes k}, V_2^{\otimes k})$ be a verifier that runs $k$ copies
of the protocol of $V$ in parallel and accepts if and only if every one of the
$k$ copies accepts.
We must show that $V'$ has completeness error at most $kc$ and soundness error
at most $s^k$.

First consider the case $x \in L$.
Let $Y=(Y_1)$ be a yes-prover that always convinces $V$ to accept with
probability at least $1 - c$.
Let $Y'=(Y_1^{\otimes k})$ be a yes-prover that runs $k$ independent copies of
the protocol of $Y$ in parallel.
Then no no-prover can win any one of the $k$ copies with probability greater
than $c$ and so by the union bound we know that the completeness error of the
repeated game is at most $kc$.

Next consider the case $x \not \in L$.
Let $N=(N_1)$ be a no-prover that always convinces $V$ to reject with
probability at least $1 - s$.
Let $N'=(N_1^{\otimes k})$ be a no-prover that runs $k$ independent copies of
the protocol of $N$ in parallel.
We now show that no yes-prover can win against $N'$ using verifier $V'$ with
probability greater than $s^k$.

Let $\Pi_{\mathrm{init}}$ denote the projection of the entire system onto the
all-$\ket{0}$ initial state.
Then the projection $\Pi_{init}' = \Pi_{\mathrm{init}}^{\otimes k}$ corresponds
to the initial state of the repeated game.
Let $\Pi_{\mathrm{acc}}$ denote the projection onto the states for which the
output qubit belonging to $V$ is 1.
Then the projection $\Pi_{\mathrm{acc}}' = \Pi_{\mathrm{acc}}^{\otimes k}$
corresponds to the accepting state of $V'$.
Let $\hilb{V}_N$ denote the Hilbert space corresponding to the private qubits
of $V$ and the private and message qubits of $N$ and let $\hilb{M}_Y$ denote
the Hilbert space corresponding to the yes-prover's message qubits.
Define
$T_N \in \mapset{T}(\hilb{V}_N \otimes \hilb{M}_Y,\hilb{M}_Y)$ as
\[
T_N(X) = \tr_{\hilb{V}_N} (\Pi_{\mathrm{init}}) X (\Pi_{\mathrm{acc}} V_2 N_1
V_1).
\]

As mentioned earlier, we may view $(V,N)$ as a new composite verifier and the
yes-prover as the lone prover for some one-message quantum interactive proof
system (i.e., a message from the prover to $(V,N)$).
In this context, Lemma 7 of Ref.~\cite{KitaevW00} asserts that the maximum
probability with which any prover could convince the verifier $(V,N)$ to
accept $x$ is precisely $\dnorm{T_N}^2.$  Because $(V,N)$ has soundness error
at most $s$, we have $\dnorm{T_N}^2 \leq s.$

Define a similar transformation $T_N' \in \mapset{T}( (\hilb{V}_N \otimes
\hilb{M}_Y)^{\otimes k}, \hilb{M}_Y^{\otimes k} )$ using $V'$, $N'$,
$\Pi_{\mathrm{init}}'$, and $\Pi_{\mathrm{acc}}'$.
It follows that $T_N' = T_N^{\otimes k}$.
{}From the multiplicativity of the diamond norm,
it follows that the maximum probability with which any prover could convince
$(V',N')$ to accept $x$ is
\[\dnorm{T_N'}^2 = \dnorm{T_N^{\otimes k} }^2 = \dnorm{T_N}^{2k} \leq s^k,\]
which establishes the desired result.

Due to the symmetric nature of quantum refereed games, we can modify the above
proof to show that $\cls{SQG}(c,s) \subseteq \cls{SQG}( c^k, ks ).$
In particular, define the verifier $V''$ so that he rejects if and only if all
$k$ copies reject.
For the case $x \not \in L$, the proof that $V''$ has soundness error $ks$
is completely symmetric to the proof that $V'$ has completeness error $kc$.

For the case $x \in L$, we let $Y$ and $Y'$ be yes-players as above.
Define the Hilbert spaces $\hilb{V}_Y$ and $\hilb{M}_N$ and the projections
$\Pi_{\mathrm{rej}}$ and $\Pi_{\mathrm{rej}}'$ in the appropriate symmetric
manner as per the above proof.
The transformation $T_Y \in \mapset{T}(\hilb{V}_Y \otimes \hilb{M}_N,
\hilb{M}_N)$ is defined as
\[
T_Y(X) = \tr_{\hilb{V}_Y} (V_1 Y_1 \Pi_{\mathrm{init}}) X (\Pi_{\mathrm{rej}}
V_2).
\]

As before, we may view $(V,Y)$ as a new composite verifier and the
no-prover as the lone prover for some quantum interactive proof system.  The
differences here are that the quantum interactive proof is now a
two-message proof instead of a one-message proof (i.e., a message from $(V,Y)$
to the prover followed by the prover's reply to $(V,Y)$) and that the prover's
goal is now to convince the verifier $(V,Y)$ to reject $x$ instead of to accept
$x$.

Fortunately, it is still straightforward to apply Lemma 7 of
Ref.~\cite{KitaevW00} to this quantum interactive proof system and so we may
claim that the maximum probability with which any prover could convince the
verifier $(V,Y)$ to reject $x$ is precisely $\dnorm{T_Y}^2.$
That $V''$ has completeness error $c^k$ follows as before.
\end{proof}

The proof of Theorem \ref{theorem:boundedError} can be extended to allow for
the slightly more general protocol wherein the verifier sends a message to the
yes-prover (via some circuit $V_{\mathrm{init}}$) before the short quantum game
commences.
This extension follows from the fact that we can apply Lemma 7 of
Ref.~\cite{KitaevW00} to the augmented transformations
\begin{eqnarray*}
T_N(X) &=& \tr_{\hilb{V}_N} (V_{\mathrm{init}} \Pi_{\mathrm{init}}) X
(\Pi_{\mathrm{acc}} V_2 N_1 V_1) ,\\
T_Y(X) &=& \tr_{\hilb{V}_Y} (V_1 Y_1 V_{\mathrm{init}} \Pi_{\mathrm{init}}) X
(\Pi_{\mathrm{rej}} V_2).
\end{eqnarray*}

Combining Theorems \ref{theorem:THEtheorem} and \ref{theorem:boundedError} we
obtain the following corollary, which is the main result of this paper.

\begin{cor}
$\cls{QIP} \subseteq \cls{SQG}$.
\end{cor}

\begin{proof}
Given a desired error bound $2^{-p}$ where $p \in \mathit{poly}$, choose
$\varepsilon \in 2^{-\mathit{poly}}$ so that $p\varepsilon \leq 2^{-p}$.
We have
$\cls{QIP} \subseteq \cls{SQG} \left( 1/2, \varepsilon \right) \subseteq
\cls{SQG} \left( 2^{-p},2^{-p} \right).$
\end{proof}

%=============================================================================%

\section{Conclusion} \label{sec:conclusion}

We introduced in this paper the quantum refereed game model of computation
and gave a short quantum game with exponentially small error for languages
with single-prover quantum interactive proof systems.
However, we have only scratched the surface of the quantum games model,
and many questions about it remain unanswered.
Some examples follow.
\begin{itemize}
\item
The two-turn game presented in this paper has an asymmetric protocol.
Is there also a two-turn quantum refereed game for $\cls{QIP}$ in which
the no-prover sends the first message, or in which the provers play
one turn in parallel?
\item
It is known that $\cls{QIP} \subseteq \cls{EXP}$.
How does $\cls{SQG}$ relate to $\cls{EXP}$?
\item
We mentioned in Section~\ref{sec:intro} that classical refereed games
characterize $\cls{EXP}$ \cite{FeigeK97}, which implies that many-turn
quantum refereed games are at least as powerful as $\cls{EXP}$.
What upper bounds can be proved on the power of refereed quantum games?
\item
We demonstrated that parallel repetition followed by a unanimous vote can
reduce error for short quantum games.
Is there a way to reduce the error in \emph{any} quantum refereed game
without affecting the number of turns in the game?
\end{itemize}

%=============================================================================%

\subsection*{Acknowledgments}

This research was supported by Canada's NSERC, the Canada Research Chairs
program, the Canadian Institute for Advanced Research (CIAR), and a graduate
student scholarship from the Province of Alberta.

%=============================================================================%

\bibliographystyle{plain}

%=============================================================================%

\end{document}